\documentclass[12pt]{iopart}
\usepackage{enumerate}

\usepackage{graphicx}

\usepackage{bm}

\def\beq{\begin{equation}}
\def\eeq{\end{equation}}

 \def\gtap{\mathrel{ \rlap{\raise 0.511ex \hbox{$>$}}{\lower 0.511ex
   \hbox{$\sim$}}}} 
\def\ltap{\mathrel{ \rlap{\raise 0.511ex
    \hbox{$<$}}{\lower 0.511ex \hbox{$\sim$}}}} 
\newcommand{\bea}{\begin{eqnarray}} \newcommand{\eea}{\end{eqnarray}}
\newcommand{\deltaunodue}{\mbox{$\Delta m_{21}^2 $}}
\newcommand{\deltasol}{\mbox{$\Delta m_{21}^2 $}}
\newcommand{\deltaunotre}{\mbox{$\Delta m_{31}^2 $}}
\newcommand{\deltaatm}{\mbox{$\Delta m_{31}^2 $}}

\begin{document}

\hfill DCPT/08/124

\hfill IPPP/08/62

\title[Sterile Neutrinos in Light of Recent Cosmological and
  Oscillation Data]{Sterile Neutrinos in Light of Recent Cosmological
  and Oscillation Data: a Multi-Flavor Scheme Approach} 
 
\author{Alessandro Melchiorri$^1$, Olga Mena$^2$, Sergio Palomares-Ruiz$^3$,
  Silvia Pascoli$^4$, Anze Slosar$^5$ and Michel Sorel$^2$} 

\address{$^1$ INFN - Sez. di Roma, Dipartimento di Fisica,
  Universit\'a di Roma "La Sapienza", I-00185 Roma, Italy} 

\address{$^2$ Instituto de F\'{\i}sica Corpuscular, IFIC, CSIC and
  Universidad de Valencia, Spain}  

\address{$^3$ Centro de F\'{\i}sica Te\'orica de Part\'{\i}culas,
  Instituto Superior Tecnico, 1049-001 Lisboa, Portugal} 

\address{$^4$ IPPP, Department of Physics, Durham University, Durham
  DH1 3LE, United Kingdom} 

\address{$^5$ Berkeley Center for Cosmological Physics, Physics
  Department and Lawrence Berkeley National Laboratory, University of
  California, Berkeley California 94720, USA}  

\begin{abstract}
  Light sterile neutrinos might mix with the active ones and be
  copiously produced in the early Universe. In the present paper, a
  detailed multi-flavor analysis of sterile neutrino production is
  performed. Making some justified approximations allows us to
  consider not only neutrino interactions with the primeval medium and 
  neutrino coherence breaking effects, but also oscillation effects
  arising from the presence of three light (mostly-active) neutrino
  states mixed with two heavier (mostly-sterile) states. First, we
  emphasize the underlying physics via an analytical description of
  sterile neutrino abundances that is valid for cases with small
  mixing between active and sterile neutrinos. Then, we study in
  detail the phenomenology of $(3+2)$ sterile neutrino models in light
  of short-baseline oscillation data, including the LSND and MiniBooNE
  results. Finally, by using the information provided by this
  analysis, we obtain the expected sterile neutrino cosmological
  abundances and then contrast them with the most recent available
  data from Cosmic Microwave Background and Large Scale Structure
  observations. We conclude that $(3+2)$ models are significantly more
  disfavored by the internal inconsistencies between sterile neutrino
  interpretations of appearance and disappearance short-baseline data
  themselves, rather than by the used cosmological data. 
\end{abstract}

\submitto{Journal of Cosmology and Astroparticle Physics}
\pacs{14.60.Pq, 98.80.-k }



\pagebreak


\section{Introduction}
\label{sec:intro}


During the last several years the physics of neutrinos has achieved
remarkable progress. The experiments with
solar~\cite{sol,SKsolar,SNO1,SNO2,SNO3,SNOsalt},
atmospheric~\cite{SKatmI,SKatmII}, reactor~\cite{KamLAND}, and also
accelerator neutrinos~\cite{K2K,MINOSprop,MINOS}, have provided
compelling evidence for the existence of neutrino oscillations,
implying non-zero neutrino masses.  

The present data require at least two large ($\theta_{12}$ and
$\theta_{23}$) angles in the neutrino mixing matrix and at least two
mass squared differences, $\Delta m_{ji}^{2} \equiv m_j^2 -m_i^2$
(where $m_{j}$'s are the neutrino masses), one driving the atmospheric
($\deltaunotre$) and the other one the solar ($\deltaunodue$) neutrino
oscillations. The mixing angles $\theta_{12}$ and $\theta_{23}$
control the solar and the atmospheric neutrino oscillations, while the
third angle $\theta_{13}$ in the three neutrino mixing matrix is bound
to be small by the data from the CHOOZ and Palo Verde reactor
experiments~\cite{CHOOZ,PaloV}. Unfortunately, oscillation experiments
only provide bounds on the neutrino mass squared differences,
i.e. they are insensitive to the overall neutrino mass scale. 

Cosmology provides one of the means to tackle the absolute
scale of neutrino masses. The Universe can be thus exploited as a new
laboratory where to test neutrino masses and neutrino physics. The new
accurate measurements of the Cosmic Microwave Background (CMB)
anisotropy and polarization from satellite, balloon-borne and
ground-based experiments have fully confirmed the predictions of the
standard cosmological model (see e.g.~\cite{WMAP5-1}) and allow us to
\emph{weigh} neutrinos. Neutrinos can indeed play a relevant role in
large scale structure formation and leave key signatures in several
cosmological data sets. More specifically, the amount of primordial
relativistic neutrinos change the epoch of the matter-radiation
equality, leaving an imprint on both CMB anisotropies (through the
so-called Integrated Sachs-Wolfe effect) and on structure formation,
while non relativistic neutrinos in the recent Universe 
suppress the growth of matter density fluctuations and galaxy
clustering (see e.g~\cite{LP06}).

All these new observations have allowed us to place new bounds on
neutrino physics from
cosmology~\cite{cosmonulimits1,cosmonulimits3,DMS06,WMAP5-1,Fogli08,nueffnew,CMSV05}
constraining the amount of energy density in neutrinos and their mass,
in the framework of three active degenerate neutrinos, to be below
$\sim0.25$~eV at $95 \%$ C.L. However, the three neutrino scenario is
a minimal scheme, inspired, following a theoretical prejudice, by a
model which tries to resemble the three family structure of quarks and
charged leptons. We now know neutrinos are different from the quark
and charged lepton sectors and there is no fundamental symmetry in
nature forcing a definite number of right-handed (sterile) neutrino
species, as those are allowed in the Standard Model fermion content.

Models with three active neutrinos plus an additional sterile one
($3+1$ models) were introduced~\cite{fournus} to explain
simultaneously solar, atmospheric and LSND~\cite{LSND} data. The mass
of a fourth, thermal, sterile neutrino has been constrained from
cosmological data to be less than $\sim0.65$~eV at $95 \%$ C.L. (see
e.g.~\cite{DMS06}). While cosmological constraints on the neutrino
mass are certainly remarkable, one should be careful in taking those
constraints blindly, since  they are obtained in a model-dependent
way. For instance, the constraint on the mass of the fourth sterile
neutrino presented in~\cite{DMS06} is obtained under the assumption of
a $(3+1)$ scenario with a fully thermalized sterile neutrino. On the
other hand, if different thermalization mechanisms are
considered~\cite{GPP04,FV95,BR92,BB01}, the cosmological constraints on
neutrino masses could be relaxed or even drastically changed.

More recently, scenarios with three active plus two sterile neutrinos
($3+2$ models)~\cite{PS01} have been shown to provide a better fit to
short baseline (SBL) data compared to $(3+1)$
models~\cite{SCS04,Karagiorgi07,MS07}, since they can, within a
neutrino oscillation framework, bring into agreement the LSND
$\bar{\nu}_e$ excess in a $\bar{\nu}_{\mu}$ beam~\cite{LSND} and the
null results from other SBL oscillation experiments.  The masses for
the additional mostly-sterile neutrinos we are interested in are much
larger than the light neutrino ones, $m_4, m_5 \gg m_1, m_2, m_3$.
Throughout this paper, we take the convention $m_5>m_4$.  The flavor
neutrino base $\nu_\alpha$, $\alpha = e, \mu, \tau, s, p$, is related
to the massive base $\nu_i$, $i=1,2,3,4,5$, through a $5\times5$
unitary matrix which we indicate as $U$:
\begin{equation}
\nu_\alpha = U_{\alpha i} \nu_i ~.
\label{eq:mixinggen}
\end{equation}
Searches of sterile neutrinos indicate that the 
mixing between heavy neutrinos and active neutrinos
and between light neutrinos and sterile ones is very small, 
so throughout our study we will take 
$|U_{a 4}|, |U_{a5}|,|U_{j s}|, |U_{jp}| \ll 1$,
with $a=e, \mu, \tau$ and $j=1,2,3$.

The way in
which these neutrinos shape the Universe depends crucially on the
thermalization processes. The purpose of the present study is to test
the cosmological viability of the $(3+2)$ scenario in which sterile
neutrinos are produced through
oscillations~\cite{BD90,EKM90a,K90,EKM90b,BD91,SMF91,EKT92,XSF93,DW94} 
(see~\cite{Dolgov81,earlynuosc,LPST87} for earlier works), applying a
multi-flavor treatment in the evolution of the system. We add in the
analysis the most recent neutrino oscillation data, which includes
the first MiniBooNE $\nu_\mu \to \nu_e$ oscillation results. Finally,
let us note that the presence of additional relativistic degrees of
freedom at the epoch of Big Bang Nucleosynthesis (BBN) modifies
significantly the production of helium, deuterium and lithium (a
recent review is given in~\cite{BBNsterile1}, see also~\cite{Dol04a}),
so this observational data on the abundance of light elements allows
one to put severe constraints on sterile neutrino parameters. However,
in the present study we consider the contribution of sterile neutrinos
to the energy density of the Universe and its effects on the CMB and
on the matter power spectrum, but we do not include the data from BBN
observations.

The structure of the paper is as follows. In Sec.~\ref{sec:formalism}
we describe the sterile neutrino production. Although we do not attempt
to solve the exact quantum kinetic equations for a $(3+2)$ system, we
have numerically computed the momentum-averaged counterpart with an
approximation for the coherence-breaking terms. This amounts to solve
an equation for the $5 \times 5$ (anti)neutrino density matrix in
order to use these solutions in our Markov chain Monte Carlo (MCMC)
analyses. We present as well an analytical approach which, under a
number of approximations, describes the neutrino thermalization in a
multi-family scenario, deriving very simple formulae that illustrate 
the evolution process in a very clear way.
Sec.~\ref{sec:data_analysis} describes the SBL oscillation data and
the cosmological data sets used in the MCMC analyses. We present our
results in Sec.~\ref{sec:results}, discussing first the compatibility
of oscillation and cosmological data and presenting afterwards a
combined analysis of both data sets. We conclude in
Sec.~\ref{sec:conclusions}.


\section{Sterile Neutrino Production in the Multi-Flavor Approach} 
\label{sec:formalism}


As a case study of sterile neutrino cosmological abundance in
multi-flavor schemes, we consider mixings among three active plus two
sterile neutrino states, i.e., the so-called $(3+2)$ sterile neutrino
models. In the Early Universe, and in particular for temperatures above
neutrino decoupling ($T > 1$~MeV), neutrinos are part of a gas of
interacting multispecies of elementary particles. In this environment,
sterile neutrinos can be produced via the oscillations of active
neutrinos in presence of interactions with the thermal plasma (for
foundational work
see~\cite{Dolgov81,damping,Stodolsky87,Manohar87,MT91,Pantaleone92,RSS93,SR93,Samuel,MT94}).Due
to the high temperatures, the neutrino mean free path is quite small
and it is essential to take into account the breaking of coherence,
using the density matrix formalism. The quantum operator for the
density matrix of neutrinos is defined as $\hat{ \rho}^i_j = \nu^i
\nu_j^\ast$. The $C$-valued density matrix, $\rho (E,T)$, is obtained
by taking the average of the matrix element over the medium. Although
the exact form of the terms which take into account the loss of
coherence is quite complicated~\cite{MT91,RSS93,SR93,MT94,BVW99}, a
good approximate treatment is obtained if the kinetic equations for
$\rho$ are written as in~\cite{Dolgovrev} (where we will omit writing
the $E$ and $T$ dependence in what follows): 
\begin{equation} \dot{\rho} = i \left[ H_{\rm m} + V_{\rm eff}, \rho
    \right]  -  \left\{ \Gamma, (\rho - \rho_{\rm eq}) \right\} ~.
\label{eq:rhoeq}
\end{equation} 
Here, $H_{\rm m} = U H_0 U^\dagger$ is the free neutrino Hamiltonian
in the flavor basis, which is obtained rotating the Hamiltonian in the
mass basis $H_0 = \mathrm{diag} (E_1, E_2, E_3, E_4, E_5)$ by the
unitary mixing matrix $U$ given by Eq.~(\ref{eq:mixinggen}). For
the temperatures under consideration, neutrinos are highly
relativistic and the approximation $E = p+ m^2/(2 p)$, with $p (\simeq
E)$ the neutrino momentum, holds and will be applied. The effective
potential $V_{\rm eff}$ describes the interactions of neutrinos with
the medium and is diagonal in the flavor basis. For negligible lepton
asymmetry, its elements are approximately given by:
\begin{equation}
V_{{\rm eff}, a} = - C_a G_F^2 T^4 E / \alpha~,
\label{eq:veff}
\end{equation}
where $G_F$ is the Fermi coupling constant, $T$ is the plasma
temperature, $E$ is the neutrino energy, $\alpha= 1/137$ the fine
structure constant. The constants $C_a$ are given by $C_e \sim 0.61$
and $C_{\mu, \tau} \sim 0.17$ (for $T< m_\mu$)~\cite{NR88}, which are
exact in the limit of thermal equilibrium of all leptons involved, and
sufficiently accurate for our purposes. The coherence loss in the
evolution of the non-diagonal terms is given by $\dot{\rho} = - \Gamma
\rho$~\cite{Dolgov81,MT91,EKT92,SR93} and a simplified but accurate 
way to describe the neutrino production and destruction is given by
the anticommutator term in Eq.~(\ref{eq:rhoeq})~\cite{Dolgovrev}. The
damping factor $\Gamma$ is diagonal in the flavor basis, $\Gamma =
\mathrm{diag}(\Gamma_e, \Gamma_\mu, \Gamma_\tau, 0,0)$, and its
elements, taking the total scattering rate, in the Boltzmann
approximation~\cite{EKT92}, are given by: 
\begin{equation}
\Gamma_a = g_a \, \frac{90 \, \zeta(3)}{7 \, \pi^4} \, G_F^2 \, T^4 \,
p~, 
\label{eq:gamma}
\end{equation}
with $a=e,\mu, \tau$. The coefficients $g_a$ have been computed
numerically and are given by $g_e\simeq 3.6$, $g_\mu= g_\tau \simeq
2.5$~\cite{DHPS00}.  The diagonal matrix $\rho_{\rm eq} = {\rm
  diag}(\rho_{eq})= I (\exp(E/T) +1)^{-1}$, with $I$ the identity
matrix, is the equilibrium value of the density matrix.

For the purpose of this work we can assume that, in the range of
temperatures under interest, the time dependence of the temperature
scales as $\dot{T} \simeq - HT$, where $H$ is the Hubble expansion
rate.  Using the fact that 
\begin{equation}
T \left( \frac{\partial \rho}{\partial T} \right)_E + E \left(
\frac{\partial \rho}{\partial E} \right)_T= 
T \left( \frac{\partial \rho}{\partial T} \right)_{\frac{E}{T}} ~,
\label{eq:tempdep}
\end{equation}
we can rewrite Eq.~(\ref{eq:rhoeq}) as
\begin{equation} 
\left(\frac{\partial \rho}{\partial T}\right)_{\frac{E}{T}} \simeq
  -\frac{1}{HT}( i\left[ H_{\rm m} + V_{\rm eff}, \rho \right]  -
  \left\{ \Gamma, (\rho - \rho_{\rm eq}) \right\}) ~,
\label{eq:rhoeq2}
\end{equation}
where 
\begin{equation}
H(T)=\sqrt{\frac{4\pi^3g_{\star}}{45}}\frac{T^2}{m_{Pl}} ~.
\label{eq:h}
\end{equation}
Here, $m_{Pl}$ is the Planck mass, while $g_\star$ is the number or
relativistic degrees of freedom. In the range of temperatures where all
the processes we are interested in take place, $g_\star$ changes very
little. Hence, we have neglected the temperature dependence of
$g_\star$ in the time dependence of the temperature, although we have
taken it into account in the expression of the Hubble rate.

The MCMC analyses presented in Sec.~\ref{sec:data_analysis} use the
solutions to the approximate evolution equation,
Eq.~(\ref{eq:rhoeq2}), for the $5 \times 5$ (anti)neutrino density
matrix $\rho$. Given the hermiticity of the $C$-valued density matrix,
this amounts to solving a set of 30 coupled first-order differential
equations. Thus, in order to reduce the computing load, to solve
Eq.~(\ref{eq:rhoeq2}), we only consider monoenergetic neutrinos at any
given temperature, taking an average value of $E=7\zeta (4)T/(2\zeta
(3))\simeq 3.15 \, T$. In general, if the oscillations are faster than
the Universe expansion rate as is the case, this is a justified
approximation~\cite{MT94}. On the other hand, we take thermal
abundances for the mostly-active neutrino triplet and zero abundances
for the mostly-sterile states, as initial conditions at a temperature
of $T \sim 100$~MeV, and follow the temperature evolution until
$T=1$~MeV~\footnote{Note that post weak decoupling effects can only be
  important when there is a large lepton number~\cite{ABFW05}.}. As a
further approximation, we assume active neutrinos keep perfect thermal
distribution throughout the evolution. This is justified as any active
neutrino population depleted due to oscillations will be immediately
repopulated thanks to its interaction with the plasma. Furthermore, if
there is no neutrino-antineutrino asymmetry as we are assuming,
active-active oscillations must have no effect at
all~\cite{LPST87,MT94}.
 
In the following, Sec.~\ref{subsec:analytical_description}, we present
an analytical description of the solutions to Eq.~(\ref{eq:rhoeq2}),
providing formulas for sterile neutrino production in the Early
Universe that are very good approximations for sterile neutrino
scenarios with small mixings among active and heavy neutrinos (and
among sterile and light neutrinos). We will use these results to show
the effects of different approximations assumed when solving
numerically the problem.


\subsection{Analytical Description}
\label{subsec:analytical_description} 
 

The production of sterile neutrinos can be computed by solving
Eq.~(\ref{eq:rhoeq}) and in particular by following the evolution of
the diagonal sterile terms $\rho_{ss}$ and $\rho_{pp}$. In the present
analytical derivation, we are interested in values of the heavy
sterile masses such that $\Delta m^2_{41}, \Delta m^2_{51} \gg
\deltaatm, \deltasol$ and small mixing between the active and heavy
sectors and between the sterile and light sectors. In this case,
sterile neutrinos never reach complete thermalization and their
abundances are much lower than the equilibrium one.

As mentioned above, the resolution of Eq.~(\ref{eq:rhoeq}) implies
solving a system of 30 coupled first-order differential equations.
However as the typical frequency for oscillations between active
($\nu_a$, $a=e,\mu,\tau$) and sterile ($\nu_h$, $h=s,p$) neutrinos,
given by
\begin{equation}
\omega_{a h}= \sqrt{ \sum_i \left( \frac{\Delta m^2_{i1} U_{ai}
  U_{hi}^\ast}{E} \right)^2  + \left( \sum_i \frac{\Delta m^2_{i1}
  (|U_{ai}|^2- |U_{hi}|^2)}{2 E}  +  V_{{\rm eff}, a} \right)^2 } ~, 
\end{equation}
is much larger than the expansion rate of the Universe $H$ in the
epoch when sterile neutrino production takes place, we can use the
static approximation~\cite{FV97,BVW99}:
\begin{eqnarray}
\dot{\rho}_{ah} = 0 ~,
\label{rhoahst} \\
\dot{\rho}_{sp} = 0 ~.
\label{rhospst} 
\end{eqnarray}
All the off-diagonal terms can be found by solving this system of 20 
coupled linear equations. Hence, this approximation reduces the
problem to the resolution of a system of 5 coupled first-order
differential equations, involving only the diagonal elements of the
density matrix, which is given by
\begin{equation} 
\dot{\rho}_{\alpha \alpha} = 2 \, \left[
  \sum_{\beta\neq\alpha} 
\mathrm{Im}\left( H_{\alpha \beta}^\ast \rho_{\alpha \beta}\right)
- \Gamma_\alpha \, \left( \rho_{\alpha \alpha} - \rho_{\rm eq} \right)
  \right]~,~~\alpha,\beta = e,\mu,\tau,s,p
\end{equation}
where we have defined:
\begin{equation}
H_{\alpha \beta} \equiv \sum_i \frac{\Delta m^2_{i 1}}{2E} U_{\alpha i}
U_{\beta i}^\ast + V_{{\rm eff}, \alpha} \delta_{\alpha \beta} ~,
\label {defH}
\end{equation}
with $V_{{\rm eff}, a}$ given by Eq.~\ref{eq:veff}, and
$V_{{\rm eff}, s}, V_{{\rm eff}, p} =0$.

For the small mixing case considered, production of sterile neutrinos
never reaches thermal abundances, i.e., $\rho_{ss}, \rho_{pp} \ll
\rho_{\rm eq}$, and the effects on the active neutrino distribution
are very small and therefore we can use the approximation $\rho_{aa}
\simeq \rho_{\rm eq}$~\footnote{Note that we will also use this
approximation for larger mixings, as those studied below. Even in that
case, the distortion of the active neutrino distribution is small
enough for our purposes and is not expected to alter substantially our
results.}.  This simplifies further the problem and, using the fact
that $H_{a \beta} \ll H_{sp}, H_{ss},H_{pp}$ holds in this case, with
$\beta = e, \mu, \tau, s,p$, the remaining three systems of four linear
equations corresponding to Eqs.~(\ref{rhoahst}), one system for each
active neutrino, can be solved to find:
\begin{eqnarray}
\rho_{as} \simeq \rho_{eq} \frac{ H_{as} \left( H_{aa} - H_{pp} + i
  \Gamma_a\right)  + H_{ap} H^\ast_{sp} }{  \left( H_{aa} - H_{ss} + i
  \Gamma_a\right)  \left( H_{aa} - H_{pp} + i \Gamma_a\right)  -
  |H_{sp}|^2} ~, 
\label{rhoas}\\ 
\rho_{ap} \simeq \rho_{eq}  \frac{ H_{ap} \left( H_{aa} - H_{ss} + i
  \Gamma_a\right)  + H_{as} H_{sp} }{  \left( H_{aa} - H_{ss} + i
  \Gamma_a\right)  \left( H_{aa} - H_{pp} + i \Gamma_a\right)  -
  |H_{sp}|^2} ~. 
\label{rhoap}
\end{eqnarray}
Although $\rho_{sp}$ turns out to be subdominant with respect to
$\rho_{as}$, it plays a relevant role in the evolution of $\rho_{ss}$
and $\rho_{pp}$ and cannot be neglected. From Eq.~(\ref{rhospst}), we
get:
\begin{equation}
\rho_{sp} \simeq \rho_{eq} \frac{  \sum_{a}  \Big(\rho_{as}^\ast
  H_{ap} - \rho_{ap} H_{as}^\ast + H_{sp} (\rho_{ss} - \rho_{pp})
  \Big) }{H_{ss} - H_{pp}}~. 
\label{rhosp}
\end{equation}

In the approximations we made, the evolution of $\rho_{ss}$ and
$\rho_{pp}$ can be explicitly given by:
\begin{eqnarray}
 \dot{\rho}_{ss} & = -
2\  \mathrm{Im} \left( \sum_a \rho_{as} H_{as}^\ast - \rho_{sp}
H_{sp}^\ast \right) ~, 
\label{rhoss} \\
\dot{\rho}_{pp} & = -
2 \ \mathrm{Im} \left( \sum_a \rho_{ap} H_{ap}^\ast + \rho_{sp}
H_{sp}^\ast \right) ~. 
\label{rhopp}
\end{eqnarray}
By substituting the expressions for $\rho_{ah}$ in Eqs.~(\ref{rhoas})
and (\ref{rhoap}) and $\rho_{sp}$ in Eq.~(\ref{rhosp}), one can obtain
the two differential equations for $\rho_{ss}$ and $\rho_{pp}$ which
can then be solved in order to find the sterile neutrino abundance in
the Early Universe. It is interesting to notice that the terms
proportional to $\rho_{ss}$ and $\rho_{pp}$ in the imaginary part in
the r.h.s. of Eqs.~(\ref{rhoss}) and (\ref{rhopp}) are real as they
are proportional to $|H_{sp}|^2$ and therefore vanish. This implies
that the system is now given by two uncoupled differential equations
which can be solved either numerical or analytically with some
approximations.

For a real mixing matrix, which will hold in the numerical study we
will perform later, the results are simplified and the evolution of
$\rho_{ss}$ and $\rho_{pp}$ depends only on the imaginary part of the
off-diagonal elements of the density matrix $\rho_{h \beta}$, $h=s,p$,
as:
\begin{equation}
\dot{\rho}_{hh} = - 2 \, \left[ \sum_{\beta\neq h}
  H_{h \beta}  \, \mathrm{Im}(\rho_{ \beta h }) \right]~,~~\beta
  = e,\mu,\tau,s,p 
\label{rhodiag}
\end{equation}
which we will use throughout the remaining of this section. For
compactness of notation, we define 
\begin{equation} 
I_{\alpha \beta} \equiv \mathrm{Im} (\rho_{\alpha \beta}) = -
  \mathrm{Im} (\rho_{\beta \alpha} ) ~.
\label{defI}
\end{equation}

From Eqs.~(\ref{rhoas}) and (\ref{rhoap}), we can find the explicit
form of $I_{as}$ and $I_{ap}$: 
\begin{eqnarray}
I_{as} & \simeq & - \left[
\frac{\left[ \left(H_{aa}-H_{pp}\right)^2 + \Gamma_a^2 + H_{sp}^2
    \right] \, H_{as}}{D} \, + \right. \nonumber
\\
& & \left. \, \frac{\left[ \left(H_{aa} - H_{ss}\right) +
    \left(H_{aa}-H_{pp}\right)\right] \, H_{sp} \, H_{ap}}{D} \,
\right] \, \Gamma_a \, \rho_{\rm eq} ~,
\label{defIas}
\\
I_{ap} & \simeq & - \left[
\frac{\left[ \left(H_{aa}-H_{ss}\right)^2 + \Gamma_a^2 + H_{sp}^2
    \right] \, H_{ap}}{D} \, + \right. \nonumber
\\
& & \left. \, \frac{\left[ \left(H_{aa} - H_{pp}\right) +
    \left(H_{aa}-H_{ss}\right)\right] \, H_{sp} \, H_{as}}{D} \,
\right] \, \Gamma_a \, \rho_{\rm eq} ~,
\label{defIap}
\end{eqnarray}
where
\begin{eqnarray}
D & \equiv & \left[\left(H_{aa} - H_{pp}\right)
    \left(H_{aa}-H_{ss}\right) - H_{sp}^2 \right]^2 + 
\\
 & & \Gamma_a^2 \, \left[ \left(H_{aa} - H_{ss}\right)^2 +
    \left(H_{aa} - H_{pp}\right)^2 + 2 \, H_{sp}^2\right] + \Gamma_a^4
\end{eqnarray}
Finally, from Eq.~(\ref{rhosp}), the element $I_{sp}$ can be further 
expressed in terms of the six $I_{ah}$,
\begin{eqnarray}
I_{sp} & \simeq & - \frac{1}{H_{ss} - H_{pp}} \, \sum_a \left(
H_{ap} I_{as} + H_{as} I_{ap} \right) ~.
\label{defIsp}
\end{eqnarray}
Note that, as already discussed for the general complex case, $I_{sp}$
turns out to be subdominant with respect to $I_{ah}$, as it is
suppressed by $H_{ah}/(H_{ss}-H_{pp})$. However, as it is clear from
Eq.~(\ref{rhodiag}), it can play a relevant role in the evolution of
$\rho_{ss}$ and $\rho_{pp}$. In that equation, whereas the prefactor
of $I_{ah}$ is $H_{ah}$, that of $I_{sp}$ is $H_{sp}$, which can
compensate the smallness of $I_{sp}$ with respect to $I_{ah}$. Hence,
whenever there is mixing between the two sterile states, $I_{sp}$ must
be taken into account. It is important to note that $I_{as}$, $I_{as}$
and $I_{as}$ are proportional to $\Gamma_a$, indicating that the
production of sterile neutrinos in the Early Universe is due to the
breaking of coherence in the evolution of active neutrinos.

Using the approximation that the time-dependence of the temperature
scales as $\dot{T} \simeq - H T$ and Eq.~(\ref{eq:tempdep}), we can
rewrite Eqs.~(\ref{rhodiag})-(\ref{defIsp}) as:
\begin{eqnarray}
H \, T \, \left(\frac{\partial \rho_{ss}}{\partial
  T}\right)_{\frac{E}{T}} \simeq 2 \sum_a H_{as} \, I_{as} - 2 H_{sp}
  \, I_{sp} ~,
\label{rhossdot} 
\\
H \, T \, \left(\frac{\partial \rho_{pp}}{\partial
  T}\right)_{\frac{E}{T}} \simeq 2 \sum_a H_{ap} \, I_{ap} + 2 H_{sp}
  \, I_{sp}  ~.
\label{rhoppdot}
\end{eqnarray}
This is the system of two uncoupled first-order differential equations
which has to be solved. Indeed obtaining the final abundance of sterile
neutrinos can be easily performed numerically.

In addition, in the case of no mixing in the sterile sector as we are
considering in this work, i.e., $U_{s5}, U_{p4} \ll 1$ ($H_{sp} \simeq
0$), Eqs.~(\ref{defIas}) and~(\ref{defIap}) further simplify (with
$h=s$ or $p$) to: 
\begin{equation}
I_{ah} \simeq - \frac{H_{ah}}{\left(H_{aa}-H_{hh}\right)^2 +
  \Gamma_a^2} \, \Gamma_a \, \rho_{\rm eq} ~,
\end{equation}
and $\rho_{hh}$ is given from Eqs.~(\ref{rhossdot})
and~(\ref{rhoppdot}) by 
\begin{equation}
\frac{\rho_{hh}}{\rho_{\rm eq}} \simeq \int_{T_{\rm dec}}^{\infty} \,
  \frac{1}{H T} \, \sum_a \,
  \frac{H_{ah}^2}{\left(H_{aa}-H_{hh}\right)^2 + \Gamma_a^2} \,
  \Gamma_a \, dT
\label{eq:rhohh}
\end{equation}
where $E/T$ is kept constant. Here, $T_{\rm dec}$ is the decoupling
temperature of the active neutrinos, which is of few MeV, depending on
flavor. The term $\Gamma_a^2$ in the denominator can be neglected as
this gives a small relative correction to the final value of the
sterile neutrino distribution of order $(\Gamma_a/V_a)^2
\sim$~$10^{-4}$. Using this, we can rewrite Eq.~(\ref{eq:rhohh}),
keeping only the dominant terms and substituting numerical factors, as 
\begin{equation}
\hspace{-2cm} \frac{\rho_{hh}}{\rho_{\rm eq}} \simeq 0.053 \, y \,
  \left(\frac{\Delta m^2_{j1}}{{\rm eV}^2}\right)^2 \,
  \frac{U_{hj}^2}{\sqrt{g_\star}} \, \sum_a \, g_a \, U_{aj}^2  \,
  \int_{x_{\rm dec}}^{\infty} \, \frac{dx}{\left(\frac{\Delta
  m^2_{j1}}{{\rm eV}^2} \, U_{hj}^2 + 3.7 \times 10^{-8} \, C_a \,
  x^2 \, y^2\right)^2}
\end{equation}
where $x \equiv (T/{\rm MeV})^3$ ($x_{\rm dec} \equiv (T_{\rm dec}/{\rm
MeV})^3$), $y \equiv E/T$ and $j = 4 \, (5)$ for $h = s \, (p)$. This
is the same integral which appears when solving the simplified
two-neutrino case~\cite{DW94}. 

A further simplification can be applied by taking into account that,
for the values of $\Delta m^2_{41}$ and $\Delta m^2_{51}$ we are
interested in ($\gtap {\cal{O}}( 1~{\rm eV}^2)$), the maximum of
production of sterile neutrinos happens at $T_{\rm max} \simeq 13~{\rm
MeV} \left(\Delta m_{j1}^2/{\rm
  eV}^2\right)^{1/6}$~\cite{BD90,K90,BD91,DW94,Dolgovrev} and thus 
$T_{\rm max}^3 \gg T_{\rm dec}^3$, so the lower limit of integration
can be safely put to 0. The relative error introduced is of order
$T_{\rm dec}^3 G_F E/T \sqrt{C_a /(\alpha \Delta m^2_{j1})} \sim
10^{-3} \sqrt{1 \ {\rm eV}^2/ \Delta m^2_{j1}}$. With these
approximations the integral can be performed analytically and the
final sterile neutrino distribution is given by:
\begin{equation}
\frac{\rho_{hh}}{\rho_{eq}} \simeq 6.6  \times 10^{-3} \,
    \sqrt{\frac{\Delta m^2_{j1}}{{\rm eV}^2}} \, \sum_a \,
    \frac{g_a}{\sqrt{C_a}} \, \left(\frac{U_{aj}}{10^{-2}}\right)^2
\label{rhossintnum} ~.
\end{equation}
where we have taken $g_\star = 10.75$ and $U_{hj} \simeq 1$. It should
be noticed that for large values of $U_{aj}$ ($\gtap 10^{-2}$), the
sterile neutrino distribution approaches the equilibrium value and our
results are not valid anymore. 

As $\rho_{hh} (E,T)$ has the same functional form as $\rho_{eq} (E,T)$,
it is straightforward to obtain the contribution to the energy density
of the Universe of each of the heavy states, which is simply given by 
\begin{equation}
\Omega_h \, h^2 \simeq 7 \times 10^{-5}  \, \left( \frac{\Delta
    m^2_{j1}}{{\rm eV}^2} \right) \, \sum_a \,
    \frac{g_a}{\sqrt{C_a}} \, \left(\frac{U_{aj}}{10^{-2}}\right)^2
\label{eq:omegahintnum}
\end{equation}

We reiterate that the results in Eqs.~(\ref{rhossintnum}) and
(\ref{eq:omegahintnum}) are only valid for $\rho_{ss}, \rho_{pp} \ll
\rho_{\rm eq}$. It is straightforward to recover the well-known
two-neutrino results from these expressions, which as one could
expect, indicate that in the case of no mixing between the sterile
states, the calculation of each of the heavy neutrino abundances
reduces to adding the individual contributions due to their mixing
with each of the active neutrinos.


\subsection{Accuracy of approximations}
\label{subsec:effect_approximations} 
 

As mentioned above, we do not attempt to solve the exact quantum
kinetic equations for a $(3+2)$ system, but instead numerically find
the momentum-averaged solutions for the simplified counterpart given
by Eq.~(\ref{eq:rhoeq2}). This approach describes the loss of
coherence in an approximate way which, along with the assumption of
monochromaticity, allows us to reduce the integro-differential system
into a system of first-order differential equations.

It is well known that the averaging over momentum is a justified
approximation if the relative oscillation phases can be ignored, as is 
the case if oscillations are not slower than the expansion rate of the
Universe~\cite{MT94}. On the other hand, it has also been
shown~\cite{Dolgovrev} that a reasonably accurate description of the
coherence loss can be achieved by making use of the anticommutator in
Eq.~(\ref{eq:rhoeq2}). 

Keeping in mind that our starting point is not the system of exact
equations but an approximate one, we discuss some further
approximations assumed in order to be able to fully solve the system
of 30 coupled first-order differential equations and study what is
their effect on the final result. In order to do so, we consider the
analytic calculation described above for the small mixing case.

\begin{figure}[t]
\centering{
\includegraphics*[width=0.7\columnwidth]{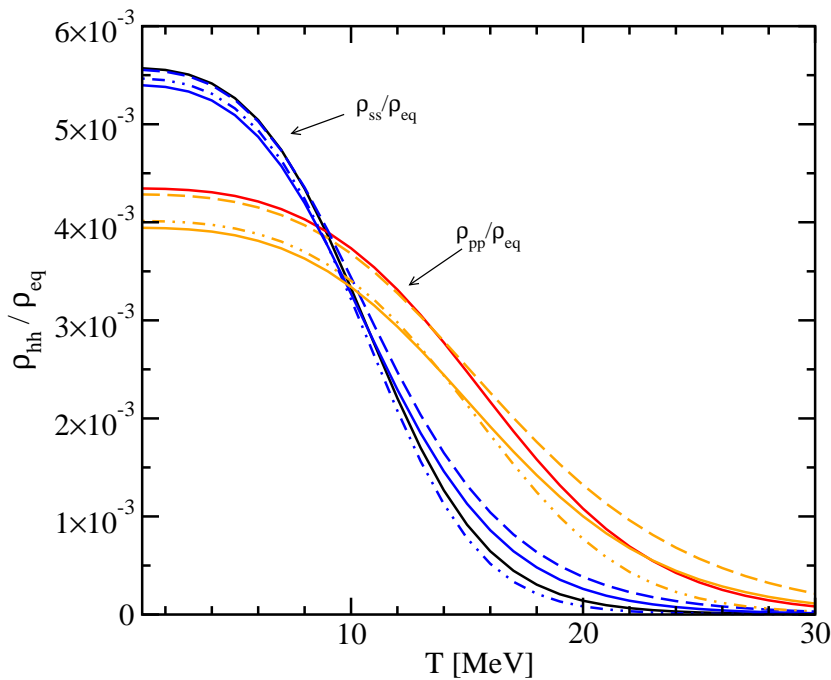}
\caption{Ratios $\rho_{ss}/\rho_{eq}$ and $\rho_{pp}/\rho_{eq}$ as a
  function of the temperature as given by Eq.~(\ref{eq:rhohh}) with
  all approximations (see text) and when only some of these
  approximations are implemented. The black (red) solid lines refer to
  the case when all approximations are taken to obtain $\rho_{ss}$
  ($\rho_{pp}$). The blue (orange) lines represent solutions to
  $\rho_{ss}$ ($\rho_{pp}$) with some approximations taken more
  accurately; dashed lines refer to the case when the momentum spread
  is taken into account, dot-dashed lines to the case  when the exact
  form of the effective potential is considered and solid lines
  represent the case when both the momentum spread and the exact
  potential are assumed. The parameters we have taken are:  $U_{e4} =
  U_{\mu 4} = 2 \times 10^{-3}$, $U_{e5} = U_{\mu 5} = 10^{-3}$, $m_4
  = 1$~eV and $m_5 = \sqrt{10}$~eV.} 
\label{fig:fig1}
}
\end{figure}

Among the different approximations used, there is the monochromaticity
of the neutrinos, assuming that active neutrinos are always in
equilibrium, neglecting the temperature dependence of the number of
degrees of freedom ($g_\star$) and using an approximate expression for
the effective potential. 

We find that the most important effect when sterile neutrinos are
close to equilibrium is due to neglecting the change of the number of
degrees of freedom as they get produced. Obviously, this approximation
has a negligible effect in the small mixing case. Surprisingly,
however, the effect of not taking into account the departure from
equilibrium of the leptons as the temperature decreases, i.e., using
an approximate expression for the effective potential, is comparable
to the previous one even in the case of small mixing. On the other 
hand, we have explicitly checked for various cases that, as expected,
solving for an averaged momentum introduces a very small
error. 

To conclude, these approximations tend to overestimate the final
abundance, although none of them induces an error larger than $\sim 10
\%$. This is depicted in Fig.~\ref{fig:fig1}, where we show the effect
of two of these approximations in the case of small mixings by solving
Eq.~(\ref{eq:rhohh}). The black (red) solid lines represent the
solution for $\rho_{ss}$ ($\rho_{pp}$) when all approximations are
implemented. By removing these approximations, we show their 
effect on the final solution. The blue (orange) lines are the
solutions to $\rho_{ss}$ ($\rho_{pp}$) when some of the approximations
are not taken. The dashed lines refer to the case when the momentum
dependence is taken into account, the dot-dashed lines represent the
case when the exact form of the effective potential is considered and
the solid lines depict the case when both the momentum spread and the
exact form of the potential are assumed. As one can see the effects
are relatively small and hence we are confident that the results we
obtain are accurate enough for our purposes. Finally, let us note that
we have also checked the perfect agreement (in this limit) between the
semi-analytic solutions of Eq.~(\ref{eq:rhohh}) and the numerical ones
of Eq.~(\ref{eq:rhoeq2}).


\section{Data Samples and Analysis Procedure}
\label{sec:data_analysis}


We now consider the signatures of the heavy mostly-sterile neutrinos
in neutrino oscillation experiments and we study the sterile neutrino
production in the Early Universe with a detailed numerical
simulation. The results of this simulation are valid both for the
$\rho_{ss}, \rho_{pp} \ll \rho_{\rm eq}$ case discussed above, as well
as for the case where sterile neutrinos reach thermal abundances
($\rho_{ss}, \rho_{pp} \sim \rho_{\rm eq}$).

\begin{figure}[t]
\centering{
\includegraphics*[width=0.47\columnwidth]{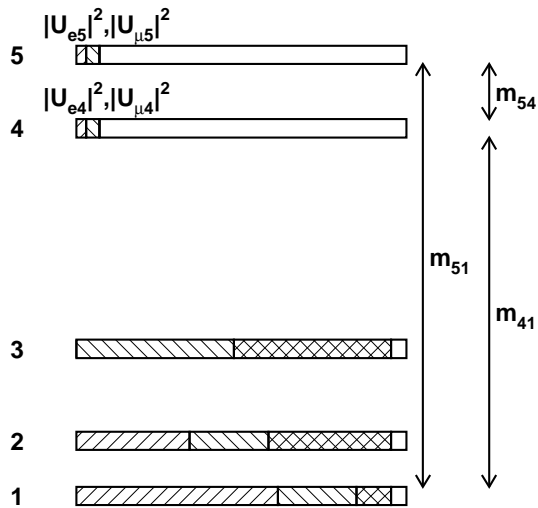}
 \caption{Flavor composition of neutrino mass eigenstates in $(3+2)$
   sterile neutrino models. The hatched rectangles indicate active
   flavor content (electron, muon, tau, respectively, from left to
   right), and the empty rectangles indicate sterile flavor content.} 
}
\label{fig:fig2}
\end{figure}

A schematic representation of $(3+2)$ models considered in this
analysis is shown in Fig.~2. We take a simplified form for the
$5\times5$ mixing matrix, requiring no mixing with the $\tau$-neutrino
sector. For the two mostly-sterile states, we allow for variable
electron and muon flavor content ($|U_{e4}|^2$, $|U_{\mu 4}|^2$,
$|U_{e5}|^2$, $|U_{\mu 5}|^2$), and variable neutrino masses $m_4$ and
$m_5$, to be constrained by our analysis described below. Concerning
the three mostly-active mass eigenstates, we fix the mass and mixing
parameters within the regions currently allowed by solar, reactor,
atmospheric, and accelerator long-baseline neutrino experiments.
Specifically, we take $m_1=0$, $m_2=\sqrt{8\times 10^{-5}}$~eV,
$m_3=\sqrt{2.5\times 10^{-3}}$~eV for the masses~\cite{GM08}, and the
upper left 3$\times$3 block form given in the following 5$\times$5
mixing matrix: 
\begin{eqnarray}
\label{eq:mixingmatrix}
U_{\alpha i}=\left( 
\begin{array}{ccccc} 
 0.81 &  0.55 & 0    & \pm |U_{e4}|    & \pm |U_{e5}| \\
-0.51 &  0.51 & 0.70 & \pm |U_{\mu 4}| & \pm |U_{\mu 5}| \\
 0.28 & -0.67 & 0.70 & 0           & 0         \\
 0    &  0    & 0    & 1           & 0 \\
 0    &  0    & 0    & 0           & 1 
\end{array}
\right) ~~,
\end{eqnarray}
where the index $\alpha = e,\mu ,\tau,s,p$ runs over flavor states,
the index $i=1,\ldots ,5$ over mass states, and we indicate the two
sterile states with the symbols $s$, $p$, respectively. The following
simplifying assumptions were therefore made: the lightest neutrino
mass eigenstate is assumed to be massless; the normal hierarchy for
the mostly-active triplet is taken; the mixing angle $\vartheta_{13}$
is taken to be zero, and $\vartheta_{12}$, $\vartheta_{23}$ are fixed
to their current best-fit values~\cite{GM08}; all six Dirac
CP-violating phases allowed for 5 neutrino species are assumed to be
zero; zero $\tau$ content in the 4th and 5th mass eigenstates is
assumed; and no mixing between the sterile states ($U_{s5} = U_{p4} = 
0$). All but maybe the first assumption are expected to have a
negligible effect; relaxing the $m_1$ condition would impose stricter
bounds on $(3+2)$ models with respect to what reported in the
following. We also note that, while do not know explicitly enforce
exact unitarity conditions in
Eq.~(\ref{eq:mixingmatrix})~\footnote{Note that not even the $3 \times 
  3$ active neutrino block is taken as exactly unitary.}, this is
approximately ensured by the smallness of the ($|U_{e4}|^2$, $|U_{\mu
  4}|^2$, $|U_{e5}|^2$, $|U_{\mu 5}|^2$) elements, providing a
sufficiently accurate description. In summary, our analysis of SBL and
cosmological data allows us to constrain six parameters of $(3+2)$
models: the masses $m_4$, $m_5$, and the mixing matrix elements
$|U_{e4}|^2$, $|U_{\mu 4}|^2$, $|U_{e5}|^2$, $|U_{\mu 5}|^2$. In the
following, combinations of these six parameters will be
shown. 

\subsection{Short-Baseline Oscillation Data}
\label{subsec:sbl_data}

Our analysis of SBL data follows closely what done
in~\cite{SCS04,Karagiorgi07}. We consider neutrino and antineutrino
data samples for three different oscillations channels:
muon-to-electron neutrino transitions, electron neutrino
disappearance, and muon neutrino disappearance. Muon-to-electron
neutrino data sets considered come from the LSND~\cite{LSND},
KARMEN2~\cite{KARMEN}, and NOMAD~\cite{NOMAD} experiments. With
respect to previous analyses, we also include recent
data~\cite{miniboonedata} on the $\nu_{\mu}\to\nu_e$ search by the
MiniBooNE experiment~\cite{MiniBooNE}. Constraints on electron
neutrino disappearance are based on data from the Bugey~\cite{Bugey}
and CHOOZ~\cite{CHOOZ} reactor experiments, while data from the
accelerator-based CDHS~\cite{CDHS} and CCFR~\cite{CCFR} experiments
were used for muon neutrino disappearance. In addition, muon neutrino
disappearance is constrained also by atmospheric~\cite{SKatmI,SKatmII}
and long-baseline~\cite{K2K} data, as done in~\cite{MSTV04}. The SBL
experimental input used is summarized in Tab.~\ref{tab:tab1}.

\begin{table}[t]
\centering{
\begin{tabular}{ccccc} \hline
Experiment & Oscillation Channel & $L/E$ (km/GeV) & Data Points \\
\hline 
KARMEN2    & $\nu_{\mu}\to\nu_e$ & 0.3-1.1        & 9  \\
LSND       &                     & 0.5-1.4        & 5  \\
MiniBooNE  &                     & 0.2-1.1        & 8  \\
NOMAD      &                     & 0.002-0.3      & 30 \\ \hline
Bugey      & $\nu_e\to\nu_e$     & 2-50           & 60 \\
CHOOZ      &                     & 100-400        & 14 \\ \hline
CCFR       & $\nu_{\mu}\to\nu_{\mu}$ & 0.004-0.03 & 18 \\
CDHS       &                         & 0.02-1.5   & 15 \\ \hline
\end{tabular}
 \caption{\label{tab:tab1} Summary of SBL data used in this analysis.}
}
\end{table}

Given the baselines and neutrino energies available to the
short-baseline experiments described above, it is possible to
constrain oscillation frequencies in the LSND allowed region, i.e., 
$\Delta m^2\sim 1$ eV$^2$, as well as the corresponding oscillation
amplitudes involving electron and muon neutrino flavors. Data are
fitted to the $(3+2)$ appearance and disappearance oscillation
probabilities applicable for short baseline experiments and described
in~\cite{SCS04,Karagiorgi07}, allowing to extract allowed regions in
the six-dimensional mass and mixing parameter space considered in this
analysis. Parameter values that best describe SBL data are given in
the top row of Tab.~\ref{tab:tab2}.

\begin{table}[t]
\centering{
\begin{tabular}{ccccccc} \hline
Data Set & $m_4$ (eV) & $m_5$ (eV) & $U_{e4}$ & $U_{\mu 4}$ & $U_{e5}$
& $U_{\mu 5}$ \\ \hline 
SBL-only    & 0.96 & 5.0  & 0.12 & 0.15 & $0.59\times 10^{-1}$ & 0.16
\\ 
SBL+cosmo   & 0.68 & 0.95 & $-0.37\times 10^{-1}$ & $0.77\times
10^{-2}$ & 0.13 & 0.19 \\ \hline 
\end{tabular}
 \caption{\label{tab:tab2} Summary of $(3+2)$ model best-fit mass and
   mixing parameters. Top row using SBL data only and bottom row for a
   combined fit of both SBL and cosmological data.} 
}
\end{table}


\subsection{Cosmological Data and Model}
\label{subsec:cosmo_data}


With the advent of the modern CMB datasets, such as those coming from
experiments like WMAP~\cite{WMAP5-2}, Boomerang~\cite{BOOM03} and
ACBAR~\cite{ACBAR} data, combined with modern galaxy surveys, for
example SDSS and 2dF surveys, cosmology has entered a precision era in
which not only basic cosmological parameters can be constrained, but
also fundamental physical parameters that enter the theory.

Cosmological neutrinos have a profound impact on cosmology since they
change the expansion history of the Universe and affect the growth of
perturbations~\cite{BS04}. Cosmological probes are sensitive to the
number of neutrinos, their relative masses and abundances. In order to
get theoretical predictions for cosmological datasets, Boltzmann codes
must be used that follow the cosmological perturbation evolution in
the linear regime, accurately accounting, in addition to the standard
cosmological constituents, for each neutrino species. In this work we
use the popular package \texttt{cosmomc}~\cite{LB02}, which has been
adapted to work with more than one massive neutrino species.

Our basic cosmological model is the minimal inflationary $\Lambda$CDM
model that is consistent with most standard cosmological probes. In
addition to neutrino parameters discussed below, its parameters are
the baryon density expressed as a fraction of the critical density
multiplied by $h^2$, $\omega_{\rm b}=\Omega_{\rm b} h^2$, where
$h=H_0/{100 {\rm km/s/Mpc}}$ is the reduced Hubble constant $H_0$,
same for the cold dark matter $\omega_{\rm dm}=\Omega_{\rm dm} h^2$,
the ratio of sound horizon to the angular diameter distance $\theta$,
which is essentially a proxy for the Hubble constant, the optical
depth to the last scattering $\tau$, logarithm of amplitude $\log A$
and spectral index of primordial fluctuations $n_s$. These basic
parameters are enough to provide a good fit to all available 
data and set other parameters. For example, the dark energy density in
a flat Universe is simply given by $\Omega_\Lambda = 1 - \Omega_{\rm
  b}- \Omega_{\rm dm}$.

The \texttt{cosmomc} code has been expanded to include 5 species 
of neutrinos. We assume
that their masses are given by 
\begin{equation}
  m_i = (0,9\times 10^{-3} {\rm eV}, 5\times 10^{-2} {\rm eV},
  m_4, m_5)
\end{equation}
and abundances by
\begin{equation}
  a_i = (1,1,1,a_4,a_5).
\end{equation}

This results in four extra parameters, $m_{4,5}$ and $a_{4,5}$, on top
of those that set the basic cosmological model. In principle, indices
4 and 5 are interchangeable and therefore likelihood space is
symmetric with respect to this change. We do not attempt to change the
parametrization to account for that, but instead use the symmetry as
an additional check for chain convergence. We use a flat prior between
zero and unity for the abundance parameter $a_i$. Since at zero
abundance, the mass can take any value, we limit the masses to be less
than 20~eV. In practice, this upper limit is not relevant for the
abundances and masses of interest. 

On the data side, we start with a conservative compendium of
cosmological datasets. These include the standard cosmic microwave 
background data, namely WMAP 5-year data~\cite{WMAP5-1,WMAP5-2},
Boomerang 03 data~\cite{BOOM03}, the latest ACBAR data~\cite{ACBAR}
and the VSA data~\cite{VSA}. In addition we use the data on the matter
power spectrum from the spectroscopic survey of Luminous Red Galaxies
(LRGs) from the SDSS survey~\cite{SDSS}. To constrain the basic
model, we also use the constraints coming from the latest compilation
of supernovae~\cite{supernovae}. Finally, we also apply the prior on
the reduced Hubble constant of $h = 0. 72 \pm 0.08$ from the Hubble key
project~\cite{HST}. 

The basic cosmological model and the basic parameter set is what we
use throughout the paper and it represents conservative constraints
from the minimal cosmological model. The check the sensitivity of our
constraints on the model used, we also calculate constraints on an
expanded model, in which we also vary the total energy density of the
Universe, or equivalently the curvature, parametrized by $\Omega_k =
1-\sum \Omega_i$, where the index $i$ runs over the various components
of the Universe and the parameter $w$ (pressure over density) that
describes the equation of state of the dark energy. In the concordance
model, these two parameters have values zero and $-1$,
respectively. On the other hand, we also use cosmological data more
aggressively and use the Lyman-$\alpha$ forest data from the SDSS
quasar sample (see e.g.~\cite{Fogli08}). This dataset provides very
good constraints on the neutrino properties and has no known
systematics at the moment, but it is an extremely difficult experiment
that has proven to be somewhat controversial.


\subsection{Analysis Procedure}
\label{subsec:analysis_procedure}

\begin{figure}[t]
\centering{
\includegraphics*[width=0.6\columnwidth]{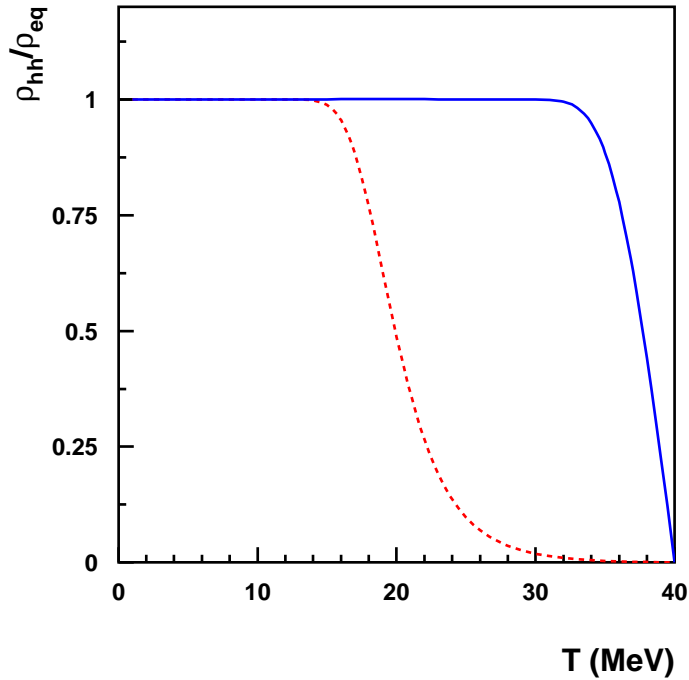}
 \caption{Evolution of sterile neutrino abundances, normalized to
   thermal equilibrium abundances $\rho_{\hbox{eq}}$, as a function of
   temperature $T$. The dashed red (solid blue) line indicate
   $\rho_{ss}/\rho_{\hbox{eq}}$ ($\rho_{pp}/\rho_{\hbox{eq}}$). The
   case shown corresponds to the $(3+2)$ model that describes SBL-only
   data best.}  
\label{fig:fig3}
}
\end{figure}

In practice, we proceed as follows. First, we generate via a MCMC
method a large number of $(3+2)$ models that give a potentially viable
description of SBL data. Specifically, in this analysis we use about
$1.6 \times 10^4$ models characterized by a SBL goodness-of-fit within
about 20~$\chi^2$ units of the SBL best-fit. Second, for each one of
those models, we determine the cosmological sterile neutrino
abundances as described in Sec.~\ref{sec:formalism}. As one example, 
we show in Fig.~\ref{fig:fig3} the behavior of the sterile neutrino 
abundances, normalized to the abundance at thermal equilibrium, as
obtained by our analysis for the $(3+2)$ model that describes SBL data
best (see top row of Tab.~\ref{tab:tab2}). Fig.~\ref{fig:fig3} shows
the evolution of $\rho_{ss}/\rho_{\mbox{}_{\mathrm{eq}}}$ and
$\rho_{pp}/\rho_{\mbox{}_{\mathrm{eq}}}$ as the temperature of the
thermal bath decreases, starting with null abundances at $T=40$~MeV
(right side of plot). For the SBL best-fit model, one can see from
Fig.~\ref{fig:fig3} that both sterile neutrino states reach thermal
abundances before decoupling. In other words, in this case, their
masses fully contribute to the sterile neutrino matter density
$(\Omega_s+\Omega_p) h^2$. Third, the viability for each of the models
considered is further assessed by contrasting the predicted sterile
neutrino abundances with cosmological observations, as explained next. 

In order to combine the short-baseline oscillation data with the
cosmological constraints, we first run extremely deep MCMC with about
70 - 100 thousands \emph{accepted} samples. Each sample is sampled from
the combined probability density in the $N$-dimensional parameter
space. The density of samples, projected to the 4-dimensional
hyper-cube of parameters ($m_4, m_5,a_4,a_5$), thus corresponds to
the probability density for those parameters, marginalized over the
remaining cosmological parameters, which are, for the sake of this
work, nuisance parameters. The relative probability of any model with
some finite 4th and 5th neutrino masses and abundances relative to the
standard minimal 3-neutrino model is calculated by comparing the
relative densities of samples at the considered model position in the
sample to the densities in the ($a_4,a_5)= (0,0)$ corner of the
parameter space. We calculate the local densities of samples in
spheres (which might be cut by the edge of the parameter space, but
this is taken into account). These are then converted to the
likelihood ratios, which are in turn converted to the effective
$\Delta \chi^2$.  This procedure allows us to importance-sample the
MCMC from the short-baseline oscillation data in
Sec.~\ref{subsec:sbl_data}.


\section{Results}
\label{sec:results}


\begin{figure}[t]
\centering{
\includegraphics*[width=0.6\columnwidth]{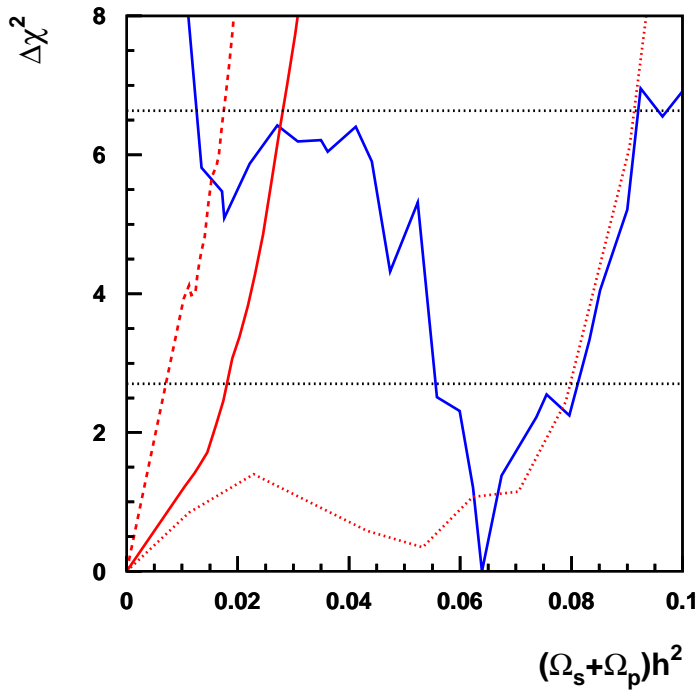}
\caption{$\Delta\chi^2$ profiles as a function of the sterile neutrino
  matter density $(\Omega_s+\Omega_p) h^2$. The red (blue) line
  indicates the case where cosmological-only (SBL-only) data are
  fitted. Three scenarios are shown for cosmological data fits:
  default data set and model (solid), data set including
  Lyman-$\alpha$ forest data and default model (dashed), default data
  set and cosmological model with free $w$ and $\Omega_k$
  (dotted). The horizontal dotted lines define the 90\% and 99\%
  confidence level regions (1 dof).}
\label{fig:fig4}
}
\end{figure}

Two types of results are presented. First, in
Sec.~\ref{subsec:results_compatibility}, we quantitatively address 
the compatibility of the SBL (including LSND and MiniBooNE) and
cosmological data sets. Second, in Sec.~\ref{subsec:results_combined},
we perform a combined analysis of SBL and cosmological data. In both
cases, the analysis is carried out under a $(3+2)$ sterile neutrino
hypothesis (see Sec.~\ref{sec:data_analysis}) and using the
multi-flavor description of cosmological sterile neutrino abundances 
(see Sec.~\ref{sec:formalism}).


\subsection{Data Sets Compatibility}
\label{subsec:results_compatibility}


\begin{figure}[t]
\centering{
\includegraphics*[width=0.6\columnwidth]{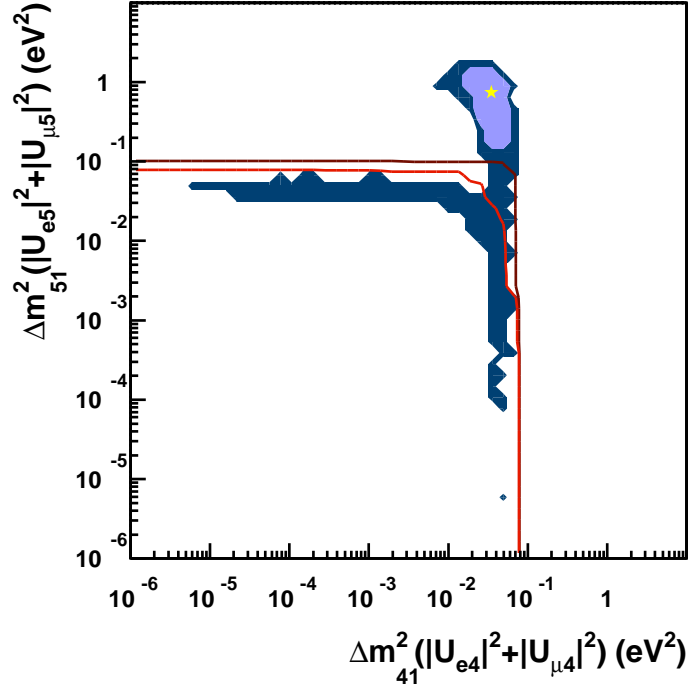}
 \caption{Allowed regions in ($\Delta m_{41}^2(|U_{e4}|^2+|U_{\mu
     4}|^2)$, $\Delta m_{51}^2(|U_{e5}|^2+|U_{\mu 5}|^2)$) space, for
     SBL-only data (filled blue regions) and cosmology-only data (red
     contours). Light (dark) colors correspond to 90\% (99\%)
     confidence level regions (2 dof). The yellow star indicates the
     SBL-only best-fit point.}
\label{fig:fig5}
}
\end{figure}

Fig.~\ref{fig:fig4} shows how the $\chi^2$ for the SBL and cosmology
data sets separately vary as a function of the sterile neutrino matter
density $(\Omega_s+\Omega_p) h^2$. The $\chi^2$ profiles for both data 
sets are shown relative to their respective best-fit $\chi^2$
values. Cosmological data prefers a small (if non-zero) value for
$(\Omega_s+\Omega_p)h^2$, while SBL data is not consistent with a null
$(\Omega_s+\Omega_p)h^2$ value, since the LSND $\bar{\nu}_e$ excess
cannot be interpreted in terms of neutrino oscillations in this
case. Quantitatively, we find the following allowed intervals, with
confidence levels for 1 degree of freedom (dof) given: 
\begin{itemize}
\item Cosmology: $(\Omega_s+\Omega_p)h^2<$0.018 (90\% CL),
  $(\Omega_s+\Omega_p)h^2<$0.028 (99\% CL)  ~,
\item SBL: 0.055$<(\Omega_s+\Omega_p)h^2<$0.081 (90\% CL),
  0.013$<(\Omega_s+\Omega_p)h^2<$0.097 (99\% CL) ~.
\end{itemize} 
In other words, no overlap is found at 90\% CL in the
$(\Omega_s+\Omega_p)h^2$ ranges allowed by the two data sets, while
overlap exists at 99\% CL. 
 
The previous numbers refer to our default cosmological data set and
default cosmological model described in Sec.~\ref{subsec:cosmo_data},
the one used as reference for all following results and figures. In
Fig.~\ref{fig:fig4}, this case corresponds to the solid red line. We
have also investigated the impact of varying the cosmological data set
and cosmological model. The dashed red line in Fig.~\ref{fig:fig4}
uses an extended data set including Lyman-$\alpha$ forest data,
providing a tighter cosmological constraint. On the other hand, the
dotted red line in Fig.~\ref{fig:fig4} quantifies by how much the
cosmological constraint is relaxed by assuming a cosmological model
with free $w$ and $\Omega_k$ (see Sec.~\ref{subsec:cosmo_data}).

\begin{figure}[t]
\centering{
\includegraphics*[width=0.6\columnwidth]{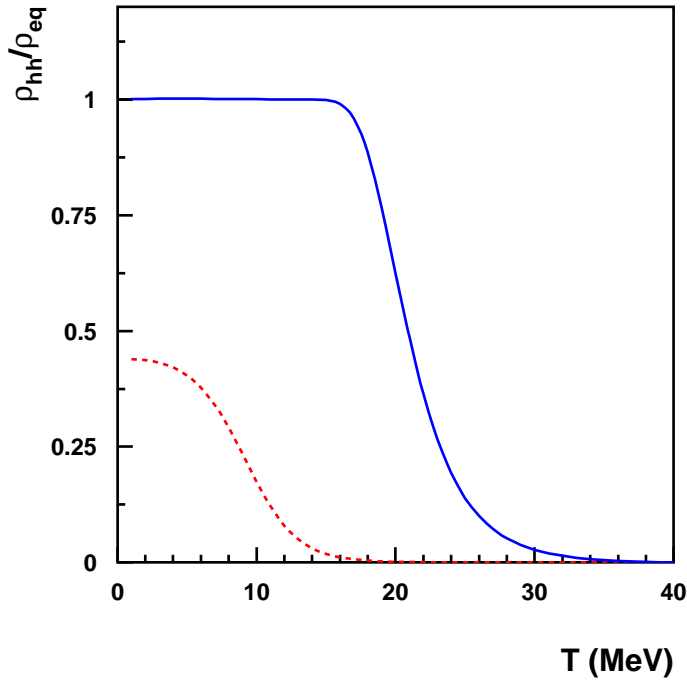}
 \caption{Evolution of sterile neutrino abundances, normalized to
   thermal equilibrium abundances $\rho_{\hbox{eq}}$, as a function of
   temperature $T$. The dashed red (solid blue) line indicate
   $\rho_{ss}/\rho_{\hbox{eq}}$ ($\rho_{pp}/\rho_{\hbox{eq}}$). The
   case shown corresponds to the $(3+2)$ model that describes all data
   (SBL plus cosmology) best.}
\label{fig:fig6}
}
\end{figure}

Fig.~\ref{fig:fig5} conveys similar compatibility information as the
one shown in Fig.~\ref{fig:fig4}, but expressed in terms of particle
physics rather than cosmological parameters. Two-dimensional allowed
regions in $(\Delta m_{41}^2(|U_{e4}|^2+|U_{\mu 4}|^2), \Delta
m_{51}^2(|U_{e5}|^2+|U_{\mu 5}|^2))$ are shown, separately for the SBL
and cosmological data sets. As the neutrino masses and/or the electron
plus muon flavor of the two heavy (mostly-sterile) states increases,
the heavy neutrino matter density $(\Omega_s+\Omega_p)h^2$ tends to
increase, either because of the larger neutrino mass or because of a
more fully thermalized abundance. Overlap is found between the regions
allowed by SBL and cosmology only at 99\%, but not at 90\%, confidence
level (2 dof). Given the cosmological constraints on
$(\Omega_s+\Omega_p)h^2$ in Fig.~\ref{fig:fig4}, we note that the
corresponding bounds on $\Delta m^2_{j1}\sum_a U_{aj}^2$, with $j=4,5$
and shown in Fig.~\ref{fig:fig5}, are in this case about two orders of
magnitude less stringent compared to what would have been expected from
Eq.~\ref{eq:omegahintnum}. As previously mentioned, the reason is that 
Eq.~\ref{eq:omegahintnum} is only applicable for $\rho_{ss}, \rho_{pp}
\ll \rho_{\rm eq}$ and is not generally valid in this case.

\begin{figure}[t]
\centering{
\includegraphics*[width=0.6\columnwidth]{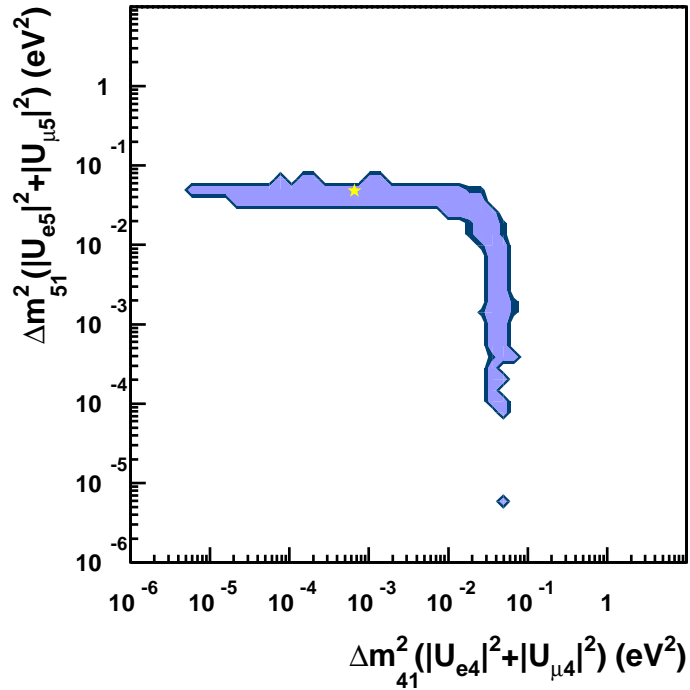}
 \caption{Allowed regions in ($\Delta m_{41}^2(|U_{e4}|^2+|U_{\mu
     4}|^2)$, $\Delta m_{51}^2(|U_{e5}|^2+|U_{\mu 5}|^2)$) space, for
     all data (SBL plus cosmology). Light (dark) blue regions
     correspond to 90\% (99\%) confidence level (2 dof). The yellow
     star indicates the global best-fit point.}
\label{fig:fig7}
}
\end{figure}

As an additional statistical test for compatibility, we consider the
``parameter goodness-of-fit'' (PG) defined in~\cite{MS03}. The test
allows to quantify how likely it is that SBL and cosmological results
arise from the same underlying $(3+2)$ sterile neutrino model. This
method alleviates a problem affecting goodness-of-fit tests based on
absolute $\chi^2$ values, namely that a possible disagreement between
the two data sets is diluted by data points which are insensitive to
the parameters that are common to both data sets. The number of
parameters common to both data sets is 4 in our case, since
cosmological data only depend on ($m_4$, $m_5$, $a_4$, $a_5$), where
the sterile neutrino abundances $a_4$ and $a_5$ defined in
Sec.~\ref{subsec:cosmo_data} are complicated functions of all mass and 
mixing parameters. This is because, in general, we cannot assume the 
sterile neutrinos to thermalize, and therefore the mixing matrix
elements (in addition to the neutrino masses) play a role in the
cosmological case also. The test is based on the statistic
$\chi^2_{\mathrm{PG}}= \chi^2_{\mathrm{PG, \ cosmo}} +
\chi^2_{\mathrm{PG, \ SBL}}$, where $ \chi^2_{\mathrm{PG, \ cosmo}}
\equiv (\chi^2_{\mathrm{cosmo}})_{\mathrm{ all \ min}}-
(\chi^2_{\mathrm{ cosmo}})_{\mathrm{ cosmo \  min}}$ and
$\chi^2_{\mathrm{ PG, \ SBL}}\equiv (\chi^2_{\mathrm{ SBL}})_{\mathrm{
    all \ min}}- (\chi^2_{\mathrm{ SBL}})_{\mathrm{ SBL \ min}}$ are
the (positive) differences for the cosmology and SBL $\chi^2$ values
obtained by minimizing the combined (cosmology plus SBL) $\chi^2$
function discussed in Sec.~\ref{subsec:results_combined}, minus the
$\chi^2$ values that best fit the two individual data sets. In our
case, we obtain $\chi^2_{ \rm PG, \ cosmo}=1.71$ and $\chi^2_{\rm PG,
  \ SBL}=5.81$, yielding $\chi^2_{\rm PG}=7.52$. The goodness of fit
parameter corresponding to this $\chi^2_{\rm  PG}$ value and four
common parameters gives a 11.1\% compatibility between the two data
sets.


\subsection{Joint Analysis}
\label{subsec:results_combined}


\begin{figure}[t]
\centering{
\includegraphics*[width=0.6\columnwidth]{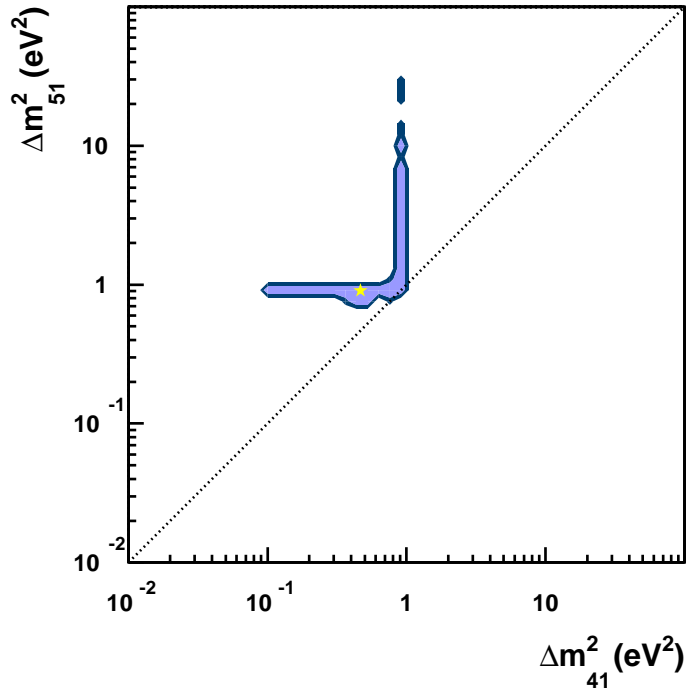}
 \caption{Allowed regions in ($\Delta m_{41}^2$, $\Delta m_{51}^2$)
   space, for all data (SBL plus cosmology). Light (dark) blue regions
   correspond to 90\% (99\%) confidence level (2 dof). The dotted line
   indicates $\Delta m_{41}^2=\Delta m_{51}^2$. The conventional
   choice $\Delta m_{41}^2>\Delta m_{51}^2$ is adopted in this
   analysis. The yellow star indicates the global best-fit point.}
\label{fig:fig8}
}
\end{figure}

Assuming that SBL and cosmological measurements arise from the same
underlying $(3+2)$ sterile neutrino model, we have also performed a
combined analysis, where the function to be minimized is the sum of
the SBL and cosmology $\chi^2$ functions. The best-fit parameters for
this combined analysis are given in the bottom row of
Tab.~\ref{tab:tab2}. As expected, the inclusion of cosmological data
tends to prefer smaller sterile neutrino masses, compared to the
SBL-only case. In addition, the best-fit mixing matrix elements are
such as to alter the thermalization of the lowest mass sterile
state. This is illustrated in Fig.~\ref{fig:fig6}, which is the
equivalent of Fig.~\ref{fig:fig3} for the best-fit SBL+cosmological
model, rather than for the SBL-only analysis. In this case, the fourth
mass state decouples with a density of only about 44\% of the thermal
one. 
 
Allowed regions in $(\Delta m_{41}^2(|U_{e4}|^2+|U_{\mu 4}|^2),
\Delta m_{51}^2(|U_{e5}|^2+|U_{\mu 5}|^2))$ space from the combined
analysis are shown in Fig.~\ref{fig:fig7}. As expected, the allowed
region in the top-right portion of Fig.~\ref{fig:fig5} disappears with
the inclusion of cosmological data in the fit, reducing the parameter
space allowed to the overlap region between the SBL and cosmological
regions.

\begin{figure}[t]
\centering{
\includegraphics*[width=0.6\columnwidth]{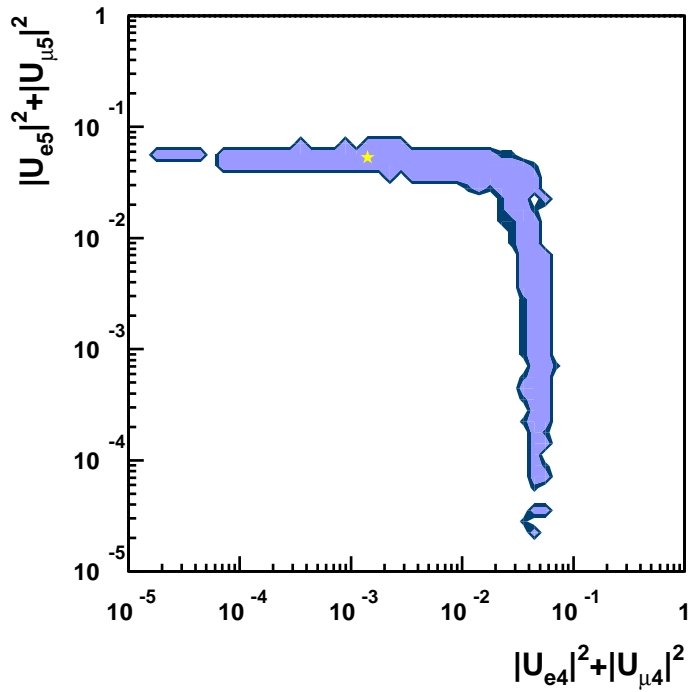}
 \caption{Allowed regions in ($|U_{e4}|^2+|U_{\mu 4}|^2$,
   $|U_{e5}|^2+|U_{\mu 5}|^2$) space, for all data (SBL plus
   cosmology). Light (dark) blue regions correspond to 90\% (99\%)
   confidence level (2 dof). The yellow star indicates the global
   best-fit point.}
\label{fig:fig9}
}
\end{figure}

Figs.~\ref{fig:fig8} and~\ref{fig:fig9} show additional
two-dimensional projections of the six-dimensional parameter space  
allowed by the combined analysis. Fig.~\ref{fig:fig8} illustrates
allowed values for the fourth and fifth mass states of $(3+2)$ models,
and Fig.~\ref{fig:fig9} their total active flavor content (electron
plus muon). In these figures, we can see that favored $(3+2)$ sterile
neutrino models are characterized by one mostly-sterile state
that is highly constrained, with $\Delta m_{j1}^2\sim$ 1~eV$^2$, 
and electron plus muon flavor content of about 5\%. On the other hand,
mass and mixing parameters for the second mostly-sterile state are far
less constrained, with mass splittings in the range $0.1$-$10$~eV$^2$,
and electron plus muon flavor content in the range $10^{-5}$-$5 \times
10^{-2}$.


\section{Conclusions}
\label{sec:conclusions}


Right-handed, sterile neutrinos can be produced in the Early Universe
via active-sterile neutrino oscillations. Depending on the structure
of the neutrino mixing matrix and the neutrino masses involved,
sterile neutrino production can be copious. In this case, sterile
neutrinos are expected to leave a clear imprint on cosmological
observables, as active neutrinos do. Given the impressive accuracy
reached by recent cosmological probes on one side, and the
well-defined nature of the cosmological concordance model capable of
interpreting those data on the other side, it is now possible to use
the Universe to constrain sterile neutrino properties that have
traditionally been probed via laboratory-based neutrino sources. 

As already extensively studied in the literature, we take into account
how both neutrino interactions with the primeval medium, and neutrino
coherence breaking effects, affect the evolution of sterile neutrino
abundances~\cite{Dolgov81,SR93,Dolgovrev}. Most previous work
generally considers only the simplest sterile neutrino mixing matrix
structures possible, that involves mixing a single sterile neutrino
species with one or more active species~\cite{CMSV05}. In this paper
we take a more general approach. We generalize those results to
account for all mixing effects arising when multiple sterile neutrino
flavors participate in the oscillations. Indeed, once sterile
neutrinos are added to the Standard Model fermion content, there is no
fundamental reason to expect only sterile single-flavor mixing
effects.

In addition to full numerical results that are valid, to a good
approximation, for generic sterile neutrino models (with the only
requirement of relativistic decoupling of sterile neutrinos), we try
to emphasize the underlying physics by developing an analytical
description that is applicable for sterile neutrinos that are
significantly more massive than the active ones, for a mixing matrix
that is approximately block-diagonal in its active and sterile
sectors, and for sub-thermal sterile neutrino abundances at
decoupling. In the latter case, we obtain the following three main
results, valid for any number of sterile neutrino species: first, that
the production of sterile neutrinos arises entirely from the breaking
of the coherence of the evolution of active neutrinos; second, that
each active flavor contributes independently to the sterile neutrino
abundances; and third, that the evolution of the abundances of
distinct sterile neutrino species is independent from one another, in
absence of mixing in the sterile sector.

Within our full multi-flavor framework, we have also studied in detail
the phenomenology of the so-called $(3+2)$ sterile neutrino
models~\cite{PS01,SCS04}. Two additional heavy (mostly-sterile)
neutrino states with masses in the eV range, beyond the minimal three
active neutrino mixing scenario, and with small amounts of electron
and muon flavor content, are considered in this case. These $(3+2)$
models were originally introduced as a means to reconcile via standard
$\bar{\nu}_{\mu}\to\bar{\nu}_e$ oscillations the currently unexplained
$\bar{\nu}_e$ excess observed by the LSND experiment with the solar 
and atmospheric oscillation signatures. We first study the sterile
neutrino evolution in the Early Universe of $(3+2)$ models that offer a
potentially viable explanation of all short-baseline oscillation data,
including the LSND excess and first $\nu_{\mu}\to\nu_e$ results
recently released by the MiniBooNE Collaboration. We then contrast the
expected energy density of the Universe in the form of sterile
neutrinos obtained by those models with the latest cosmological
observations coming from cosmic microwave background datasets and
galaxy surveys, mostly (but not only) in the framework of the minimal
$\Lambda$CDM cosmological model.  

In contrast with some literature, we find that fully solving the
neutrino kinetic equations is necessary even for the relatively
large active-sterile mixing implied by this class of models, since
sterile neutrino states do not always feature thermal abundances at
decoupling. This result, valid for the minimal $\Lambda$CDM
cosmological model, should be taken into account in future studies on
sterile neutrino production in the Early Universe, in order to avoid
possibly misleading conclusions. We find that the $(3+2)$ model that
best describes short-baseline oscillation data is excluded at high
confidence level by cosmological observations, which do not allow for
large amounts of energy density in the form of relativistic species in
addition to the energy density of active neutrinos, unless non-minimal
$\Lambda$CDM models are used. Nevertheless, by fully exploring the
neutrino parameter space, we do find $(3+2)$ sterile neutrino models
that provide a perfectly acceptable description of short-baseline and
cosmological data simultaneously, with a 11\% probability for
compatibility between the two datasets. As a consequence, we conclude
that $(3+2)$ models are significantly more disfavored by the internal
inconsistencies between sterile neutrino interpretations of appearance
and disappearance short-baseline data themselves~\cite{MS07}, rather
than by the used cosmological data. Finally, a global analysis of all
short-baseline plus cosmological data allow us to further constrain
the $(3+2)$ neutrino parameter space, in a region that should be
accessible by future experiments aiming to study neutrinos, both the
ones permeating the cosmos and neutrinos produced on Earth.


\ack
OM is supported by a \emph{Ram\'on y Cajal} contract from the Spanish
Government. SPR is supported by the Portuguese FCT through the
projects POCI/FP/81919/2007 and CFTP-FCT UNIT 777, which are partially
funded through POCTI (FEDER). SPR is also partially supported by the
Spanish Grant FPA2005-01678 of the MCT. MS would like to acknowledge
support by the Spanish Ministry of Science and Innovation via a CSIC
JAE-DOC contract, and use of the computing cluster of the experimental
neutrino group at IFIC for this work.

\end{document}